\documentclass[conference]{IEEEtran}
\IEEEoverridecommandlockouts

\usepackage{cite}
\usepackage{amsmath,amssymb,amsfonts}
\usepackage{algorithmic}
\usepackage{graphicx}
\usepackage{textcomp}
\usepackage{xcolor}

\usepackage{hyperref}
\usepackage{amsmath}
\usepackage{amssymb}
\usepackage{graphicx} 
\usepackage{subfigure} 
\usepackage{color}
\usepackage{url}
\usepackage{caption} 
\usepackage{multirow}
\usepackage{booktabs}
\usepackage{subcaption}
\usepackage{color}
\usepackage{bm}
\usepackage{amsfonts}
\usepackage{makecell}
\usepackage{xcolor}
\usepackage{newpxtext}
\usepackage{array}
\usepackage{cite}
\usepackage{bbding}
\usepackage{hyperref}

\newcommand{\etal}{\emph{et al.}}

\def\BibTeX{{\rm B\kern-.05em{\sc i\kern-.025em b}\kern-.08em
    T\kern-.1667em\lower.7ex\hbox{E}\kern-.125emX}}
\begin{document}

\title{Semantic Ensemble Loss and Latent Refinement for High-Fidelity Neural Image Compression\\

\thanks{* Corresponding Author: Yuanchao Bai}
}


\author{
\IEEEauthorblockN{
Daxin Li, 
Yuanchao Bai*, 
Kai Wang,
Junjun Jiang,
Xianming Liu
}
\IEEEauthorblockA{
\textit{Faculty of Computing, Harbin Institute of Technology}\\
hahalidaxin@stu.hit.edu.cn, yuanchao.bai@hit.edu.cn
}
}
\maketitle

\begin{abstract}
Recent advancements in neural compression have surpassed traditional codecs in PSNR and MS-SSIM measurements. However, at low bit-rates, these methods can introduce visually displeasing artifacts, such as blurring, color shifting, and texture loss, thereby compromising perceptual quality of images. To address these issues, this study presents an enhanced neural compression method designed for optimal visual fidelity. We have trained our model with a sophisticated semantic ensemble loss, integrating Charbonnier loss, perceptual loss, style loss, and a non-binary adversarial loss, to enhance the perceptual quality of image reconstructions. Additionally, we have implemented a latent refinement process to generate content-aware latent codes. These codes adhere to bit-rate constraints, and prioritize bit allocation to regions of greater importance. Our empirical findings demonstrate that this approach significantly improves the statistical fidelity of neural image compression. 
\end{abstract}

\begin{IEEEkeywords}
Generative Image Compression, Learned Image Compression
\end{IEEEkeywords}

\section{Introduction}
\label{sec:intro}

Powered by deep learning, Learned Image Compression (LIC) has undergone remarkable progress in recent times, as evidenced by numerous studies~\cite{bai2021learninga, bai2024deep, li2024groupedmixer,wang2023learning, wang2024learninglosslesscompressionhigh,he2022elic}. Some cutting-edge learning-based approaches now outperform traditional codecs like VVC~\cite{bross2021overview}, marking a promising future for LIC technologies. However, a notable challenge arises when these models, optimized for Mean Squared Error (MSE) metrics, produce unsatisfactory visual artifacts at lower bitrates. These artifacts, manifesting as blur and lost textures, result from a misalignment between the optimization objectives and the complexities of human visual perception. Additionally, human observers tend to focus on regions of higher salience or relevance in an image, such as faces, distinct objects, or text, while overlook less critical backgrounds or blurred sections. Thus, uniform bit allocation across an image can adversely affect the perceptual quality of salient areas, especially under limited bandwidth.

Previous methods have addressed image compression challenges by incorporating perceptual and adversarial losses~\cite{agustsson2019generativea,mentzer2020high,gao2021perceptual, he2022b, liang2024image}. In this work, we enhance the statistical fidelity of reconstructed images by proposing a semantic ensemble loss and latent refinement. 
The contributions are summarized as follows:
\begin{itemize}
\item We propose a semantic ensemble loss that includes Charbonnier loss, perceptual loss, style loss, and a non-binary adversarial loss. This loss function enables the model to generate more detailed and content-rich images with higher fidelity.
\item We refine the latent representations using semantic ensemble loss coupled with stochastic Gumbel annealing. This process can extract latent codes that comply with bit-rate constraints, and allocate more bits to salient image regions.
\item Extensive experiments demonstrate that our codec achieves superior perceptual quality compared to existing methods, as evaluated on the CLIC 2024 validation dataset.
\end{itemize}

\section{Method}
\label{sec:method}

\begin{figure*}[h]
    \centering
    \includegraphics[scale=0.9]{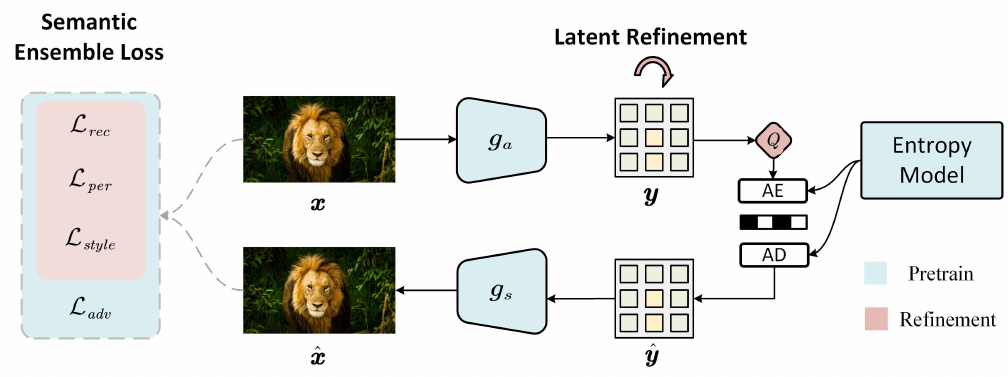}
    \caption{
    Framework Overview: Our process begins with training a VAE-based compression model employing our semantic ensemble loss, which is subsequently refined using SGA and a computationally simplified loss.
    } 
    \label{fig:overview}
\end{figure*}

\subsection{Architecture}
The overview of our framework is shown in Fig.~\ref{fig:overview}.
Our methodology follows the architecture established by HiFiC ~\cite{mentzer2020high}, which includes analysis, synthesis, hyper-analysis, and hyper-synthesis transforms. To improve entropy prediction and efficiency, we incorporate the efficient spatial-channel context model from ELIC \cite{he2022elic}. This model segments the latent representations into 10 groups and applies autoregressive prediction to each group.

\subsection{Training with Semantic Ensemble Loss}

We train a lossy compression model guided by a rate-distortion loss function, a methodology validated as effective by ~\cite{mentzer2020high, he2022b, muckley2023}, and articulated as:
\begin{equation}
\mathcal{L} = \lambda^* \mathcal{R} + \mathcal{D}
\end{equation}
Here, $\mathcal{R}$ and $\mathcal{D}$ represent the rate and distortion contributions, respectively. The hyper-parameter $\lambda^*$ modulates the balance between these losses, imposing a penalty on the rate if it exceeds the target rate $\tau$:
\begin{equation}
\label{eq:lambda}
\lambda^* = \begin{cases}
\lambda_a, & \text{if } \mathcal{R}^* \leq \tau \\
\lambda_b, & \text{otherwise}
\end{cases}, \text{with $\lambda_b \gg \lambda_a$}.
\end{equation}
Note that $\mathcal{R}^*$ is the actual rate, derived from the quantized latents, whereas $\mathcal{R}$ is computed from the noised latents to enable differentiability.
The distortion is quantified using our proposed semantic ensemble loss that incorporates reconstruction, perceptual, style, and adversarial losses:
\begin{equation} \label{eq:distortion_losses}
\mathcal{D} = \alpha \mathcal{L}_{\text{rec}} + \beta \mathcal{L}_{\text{per}} + \gamma \mathcal{L}_{\text{style}} + \delta \mathcal{L}_{\text{adv}}.
\end{equation}
In this equation, $\alpha$, $\beta$, $\gamma$, and $\delta$ are hyper-parameters that balance the four respective loss components. The reconstruction loss, $\mathcal{L}_{\text{rec}}$, employs the Charbonnier loss\cite{lai2017deep} to measure fidelity between the original and reconstructed images. The perceptual loss~\cite{zhang2018unreasonable}, $\mathcal{L}_{\text{per}}$, is derived from the $L_2$ loss between feature representations extracted by VGG of the original and reconstructed images. The style loss~\cite{sajjadi2017enhancenet, he2022b}, $\mathcal{L}_{\text{style}}$, calculates the $L_2$ loss between the Gram matrices of $16 \times 16$ feature patches from the original and reconstructed images, emphasizing differences in local textures. Lastly, the adversarial loss, $\mathcal{L}_{\text{adv}}$, is associated with the discriminator's loss. 

To further enhance the visual quality of the reconstructed images, we have integrated a non-binary adversarial discriminator within the Vector Quantized-Variational AutoEncoder (VQ-VAE) framework, following the approach by Muckley \etal~\cite{muckley2023}. This discriminator significantly improves the decoder's ability to distinguish between real and reconstructed images, focusing on capturing differences at a more refined semantic level. The adversarial loss associated with this discriminator is expressed as:

\begin{equation}
\begin{aligned}
\mathcal{L}_{\text{disc}}(\bm{\phi}) &=  \mathbb{E}_{\bm{x} \sim P_{\bm{X}}}\left[-\left\langle u(\bm{x}), \log D_{\bm{\phi}}(\bm{x})\right\rangle\right] \\
& +\mathbb{E}_{\hat{\bm{x}} \sim P{\hat{\bm{X}}}}\left[-\left\langle \bm{b}_0, \log D_{\bm{\phi}}(\hat{\bm{x}})\right\rangle\right], \\
\mathcal{L}_{\text{adv}}(\bm{\varphi}, \bm{\omega}, \bm{v}) &= \mathbb{E}_{\hat{\bm{x}} \sim P_{\hat{\bm{X}}}}\left[-\left\langle u(\bm{x}), \log D_{\bm{\phi}}(\hat{\bm{x}})\right\rangle\right],
\end{aligned}
\end{equation}
where $D_{\bm{\phi}}$, the discriminator, is parameterized by $\bm{\phi}$. Here, $u(\bm{x})$ is a one-hot vector corresponding to the index of the nearest neighbor in the codebook, and $\bm{b}_0$ denotes the absence of such a neighbor (the 'fake' label). The parameters $\bm{\varphi}$, $\bm{\omega}$, and $\bm{v}$ represent the encoder, the entropy model, and the decoder, respectively.

\begin{figure*}[!ht]
\centering     
\subfigure[]{\label{fig:d}\includegraphics[width=0.31\textwidth]{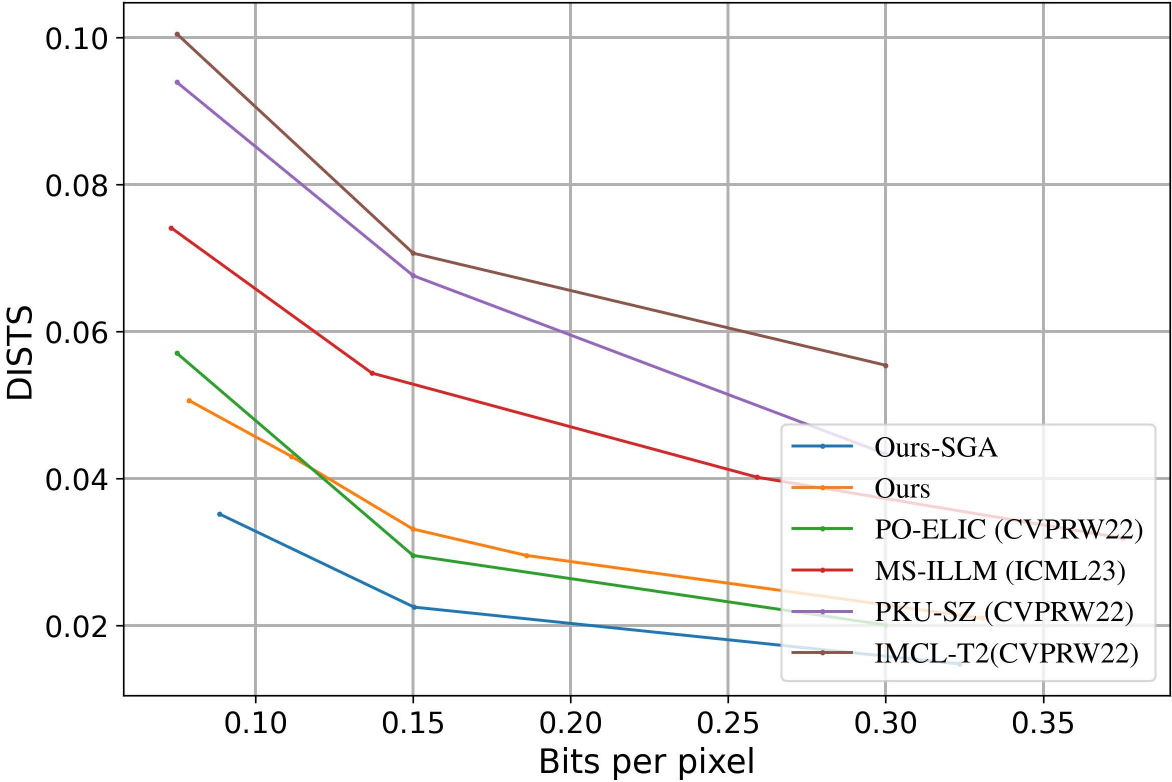}}
\subfigure[]{\label{fig:e}\includegraphics[width=0.31\textwidth]{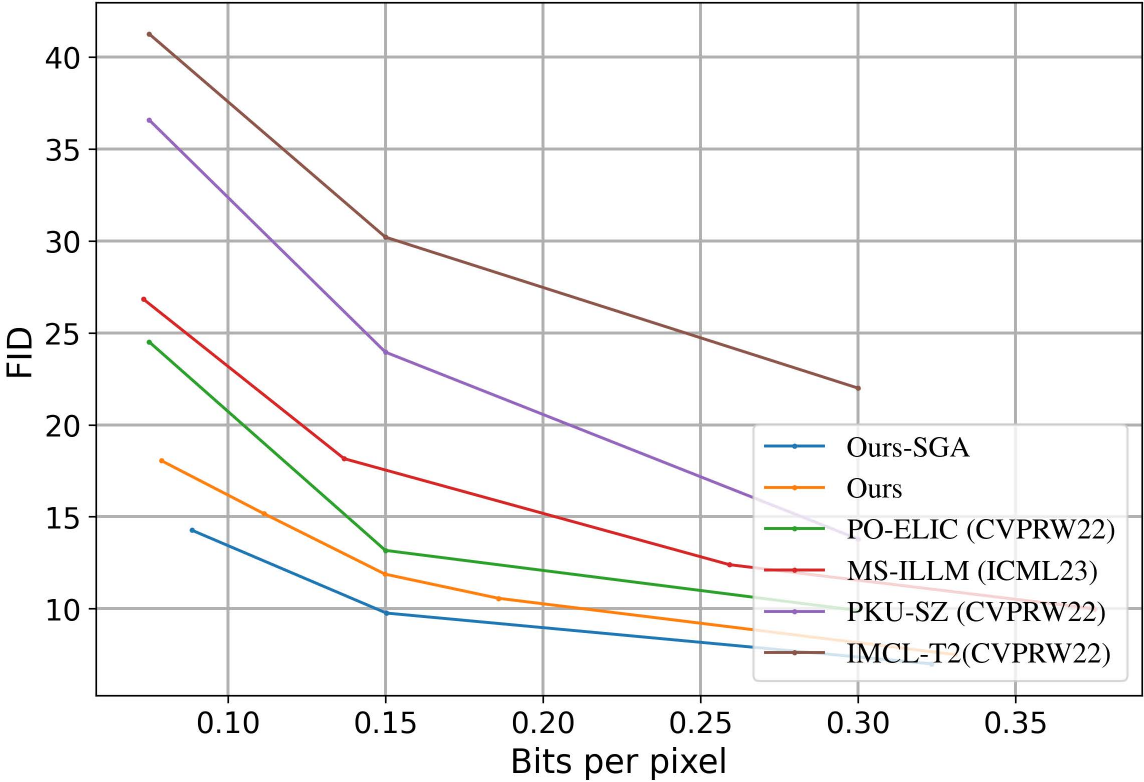}}
\subfigure[]{\label{fig:f}\includegraphics[width=0.31\textwidth]{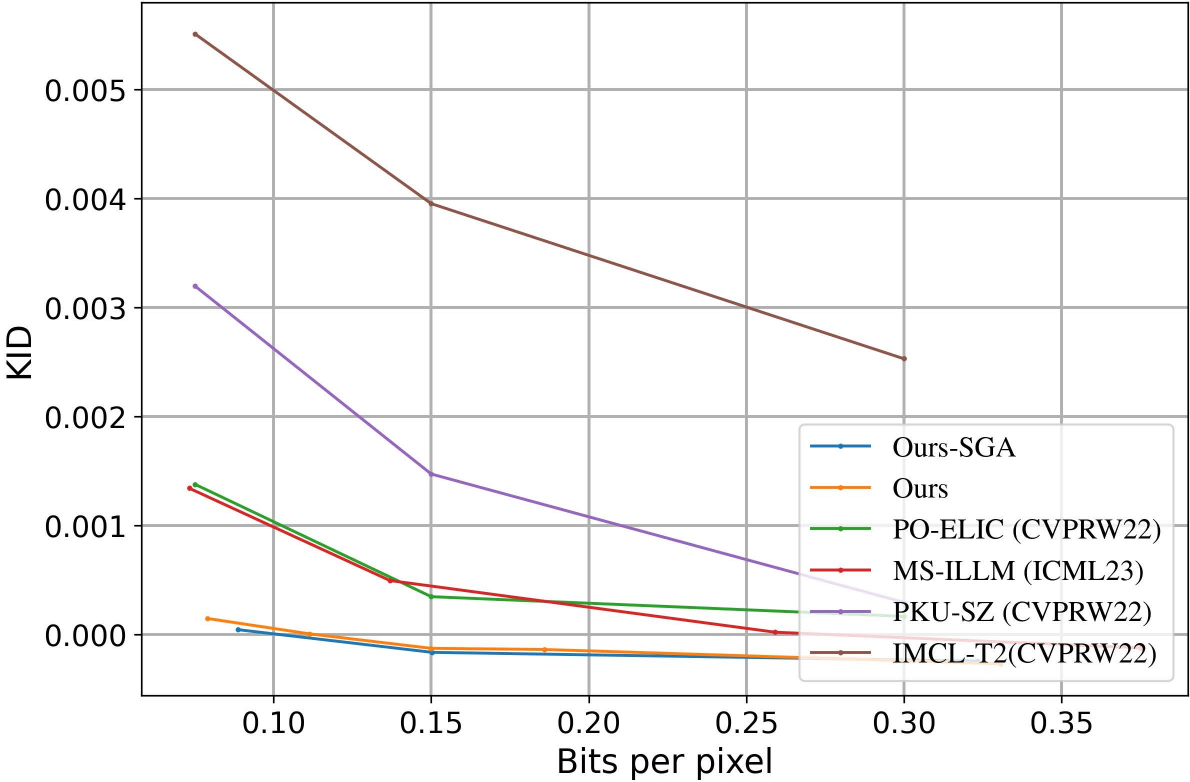}}
\caption{Comparisons of methods across various distortion and statistical fidelity metrics for the CLIC 2024 validation set. }
\label{fig:rate_plots}
\end{figure*}

\subsection{Latent Refinement for Perception}

To enhance alignment with the original image content, we apply semi-amortized optimization to derive content-adaptive latent representations. Drawing on the methods of Yang \etal\cite{yang2020improving} and Gao \etal\cite{gaoFlexibleNeuralImage2022}, we employ Stochastic Gumbel Annealing (SGA) to generate the modified latent variable, and retrieve the dequantized latent representation $\tilde{y}$ through a learnable process as described in Gao \etal\cite{gaoFlexibleNeuralImage2022}:
\begin{equation}
\tilde{\bm{y}} = \text{SGA}( \bm{y} / \Delta ) \Delta,
\end{equation}
where $\Delta$ is the learnable quantization step. For hyper latnet variable $z$, we only use SGA without learnable quantization to derive $\bar{z}$.
We manage the bitrate with a rate-constrained loss, including a simplified semantic ensemble loss as distortion, defined as follows:

\begin{equation}
\label{eq:sga_loss}
\begin{aligned}
\mathcal{L}(\bm{y}, \bm{z}) &= \lambda^* \cdot \mathcal{R} + \mathcal{D} \\
&= - \lambda^* \cdot (\log p_{\bm{\omega}}(\tilde{\bm{y}};\bar{\bm{z}}) + \log p_{\bm{\omega}}(\bar{\bm{z}})) \\&+ \alpha \mathcal{L}_{\text{rec}} + \beta \mathcal{L}_{\text{per}} + \gamma \mathcal{L}_{\text{style}},
\end{aligned}
\end{equation}
where $\lambda^*$, derived from Eq.~\ref{eq:lambda}, ensures the bitrate aligns with the target. The hyperparameters $\alpha$, $\beta$, and $\gamma$ balance the trade-off between distortion and fidelity. The adversarial loss is excluded due to its intensive computation.
Furthermore, as demonstrated in Gao \etal\cite{gaoFlexibleNeuralImage2022}, by refining the latent representations, we can redistribute bits to prioritize more significant regions. Specifically, we compute distortion losses for foreground and background regions separately, leveraging image-wise calculations for perceptual and style losses. This process is encapsulated by:
\begin{equation}
\mathcal{D}_{\text{roi}} = \lambda_{\text{fg}}\cdot \bm{m} \odot \mathcal{D}_{\text{fg}} + \lambda_{\text{bg}} \cdot (1-\bm{m}) \odot \mathcal{D}_{\text{bg}},
\end{equation}
where $\bm{m}$ denotes the foreground mask. The hyperparameters $\lambda_{\text{fg}}$ and $\lambda_{\text{bg}}$ balance the foreground and background losses, $\mathcal{D}_{\text{fg}}$ and $\mathcal{D}_{\text{bg}}$, respectively, which include similar terms to those in Eq.~\ref{eq:sga_loss}.
To obtain a dedicated mask for refining the latent representations, we utilize label tools\footnote{https://github.com/anuragxel/salt} assisted by Segment Anything (SAM)\cite{kirillovSegmentAnything2023}. SAM conveniently accepts mouse points as input and generates a corresponding mask with finer edges, as illustrated in Fig.~\ref{fig:roi_compare}.

\section{Experiments}
\label{sec:exp}

\subsection{Settings}

Our model is trained using $256 \times 256$ patches extracted from the test split of the OpenImages V7 dataset~\cite{krasin2016openimages}. We employ the AdamW optimizer with parameters $\beta_1= 0.9$ and $\beta_2= 0.999$, conducting the training over 2 million iterations. Our training methodology is in line with that of Muckley \etal~\cite{muckley2023}, which involves a two-stage process for generative models. Initially, we perform end-to-end training with a rate-constrained loss function, optimizing both rate and distortion losses, including reconstruction, perceptual, and style loss components. This is followed by a fine-tuning stage of the decoder to incorporate an adversarial loss using a non-binary discriminator. The hyperparameters in Eq.~\ref{eq:distortion_losses} are experimentally set to $\alpha=10, \beta=1, \gamma=80, \delta=0.008$ respectively during training.

For benchmarking, we compare our results with several state-of-the-art models, including MS-ILLM~\cite{muckley2023}, and leading entries from the CLIC2022 challenge, such as PO-ELIC~\cite{he2022b}, PKU-SZ~\cite{ma2021}, and IMCL-T2~\cite{Pan2022a}. We obtain reconstructed images from the official CLIC2022 challenge website and calculate metrics following the methodology described in Muckley \etal~\cite{muckley2023}.
For evaluation, we use the validation set of CLIC2024 (test set of CLIC2022), comprising 30 high-resolution 2k images. Performance is assessed using reference metrics—such as DISTS~\cite{ding2020} (better than PSNR, MS-SSIM in measuring visual consistency)—and non-reference metrics, including FID~\cite{heusel2017gans} and KID~\cite{sutherland2018demystifying}. 

\begin{figure*}[!t]
\begin{center}
 \includegraphics[scale=1.35]{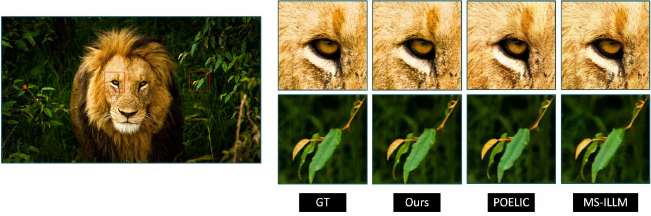}
\end{center}
\caption{\label{fig:visual_comparison}%
Visual comparisons using different methods at the same bitrate.
}
\end{figure*}
\begin{figure*}[!t]
\begin{center}
 \includegraphics[scale=0.7]{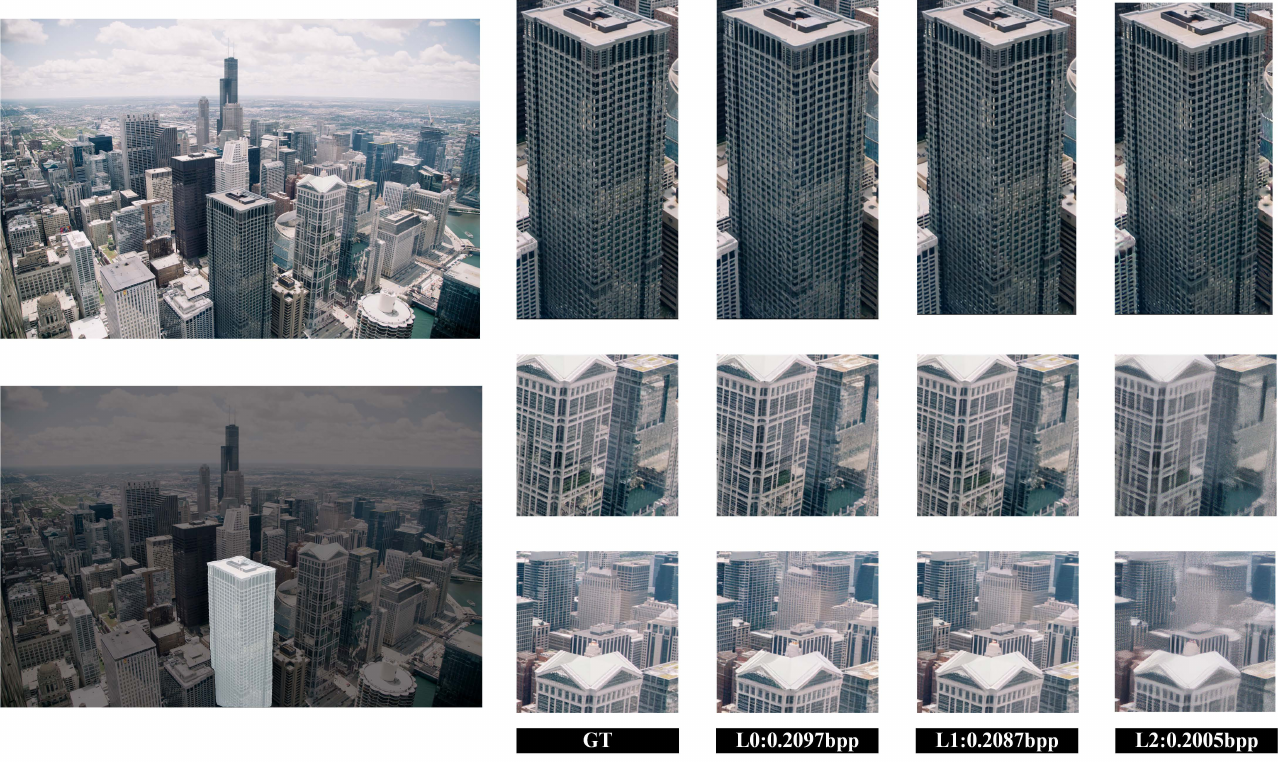}
\end{center}
\caption{\label{fig:roi_compare}%
Qualitative comparison of refining latent representations using different sets of hyperparameters for ROI-based loss.
}
\end{figure*}

\subsection{Main Performance}

Our method exhibits superior performance in FID and KID metrics, outshining all compared methodologies as referenced in MS-ILLM ~\cite{muckley2023}, PO-ELIC ~\cite{he2022b}, PKU-SZ \cite{ma2021}, and IMCL-T2 \cite{Pan2022a}. In the DISTS metric, it matches the top performance of PO-ELIC\cite{he2022b}. Specifically, our approach surpasses the former state-of-the-art, MS-ILLM\cite{muckley2023}, with a remarkable 42.49\% BD-Rate reduction in the FID metric, a 58.87\% reduction in the KID metric, and a 53.00\% reduction in the DISTS metric. The Ours-SGA variant of our method, which further refines latent representations, notably enhances both statistical fidelity and textural details. Consequently, Ours-SGA achieves the highest scores across FID, KID, and DISTS metrics. Remarkably, in comparison with MS-ILLM~\cite{muckley2023}, Ours-SGA attains a more substantial BD-Rate reduction of 62.00\% in the FID metric, 67.06\% in the KID metric, and a significant 74.63\% in the DISTS metric, underscoring our methodology's effectiveness in producing images of superior perceptual quality.

\subsection{Qualitative Results}
As shown in Fig.~\ref{fig:visual_comparison}, at the same bitrate, visual comparisons of the compression and reconstruction images using different methods demonstrate that our approach excels in generating details and maintaining consistency with the original image. Specifically, it outperforms POELIC and MS-ILLM in preserving finer details, such as the texture around the lion's eye and the leaf's structure, achieving higher fidelity to the original content.

Fig.~\ref{fig:roi_compare} displays visual results that validate the effectiveness of ROI-based latent refinement. The left two images show the original image and its corresponding mask. For comparison, we include results from $L_0$ training without ROI-based loss and from methods $L_1$ and $L_2$, employing ROI-based loss with different weights $\lambda_{fg}=(1, 1)$ and $\lambda_{bg}=(0, 0.2)$. ROI-based loss enhances fidelity and texture details in the foreground, while training focused solely on ROI distortion leads to noticeable artifacts, as seen in $L2$. The distinct weighting of foreground and background in $L1$ yields optimal performance, enhancing foreground quality while preserving background regions, demonstrating our method's efficacy in bit allocation to enhance perceptual quality.

\subsection{Ablation Study}

Our ablation studies aimed to assess the impact of different components in the proposed semantic ensemble loss. We trained models with varying losses at 0.15 bpp and evaluated their performance. The results, presented in Tab.~\ref{tab:ablation_study}, indicate that omitting style loss, perceptual loss, and adversarial loss results in FIDs of 12.44, 13.32, and 13.85, respectively. These findings highlight the significance of each component in the semantic ensemble loss, particularly the non-binary adversarial loss, which proves to be the most influential.
\begin{table}[!t]
\centering
\caption{\label{tab:ablation_study}
    Ablation studies on different terms in semantic ensemble loss.
    }
\scalebox{1}{
\begin{tabular}{l c}
\toprule
Methods   & FID↓ \\ \midrule
Ours  & 11.88   \\
w/o style loss  &  12.95  \\ 
w/o perceptual loss  &  13.44  \\ 
w/o non-binary adversarial loss  &  13.85  \\ 
\bottomrule
\end{tabular}\label{subjective_result}
}
\end{table}

\section{Conclusion}
\label{sec:conclusion}

This study introduces a codec optimized for visual quality, leveraging semantic ensemble loss and latent representation refinement. Our codec integrates semantic ensemble loss to enhance the model's ability to generate finer details. To achieve better performance and bit allocation, we employed SGA with learnable quantization for fine-tuning latent representations. Extensive experiments demonstrate that our method sets a new benchmark in image compression, excelling in perceptual quality compared to prior works.

\section*{Acknowledgment}
This work was supported in part by National Key Research and Development Program of China under Grant 2022YFF1202104, in part by National Natural Science Foundation of China under Grants 62301188, 92270116 and U23B2009, in part by China Postdoctoral Science Foundation under Grant 2022M710958, and in part by Heilongjiang Postdoctoral Science Foundation under Grant LBH-Z22156.

\bibliographystyle{IEEEtranS}
\bibliography{refs}

\end{document}